# 'CRAFT @ Large': Building Community Through Co-Making

Yiran Zhao[1], Maria Alinea-Bravo[2], and Niti Parikh[3]

[1]Yiran Zhao; Information Science, Cornell Tech; email: yiran.zhao@info.cornell.edu
[2]Maria Alinea-Bravo; NYC H+H/Coler on Roosevelt Island; email:maria.alineabravo@nychhc.org
[3] Niti Parikh; MakerLAB, Cornell Tech; email: ntp27@cornell.edu

## INTRODUCTION

'CRAFT @ Large' (C@L) is an initiative launched by the MakerLAB at Cornell Tech to create an inclusive environment for the intercultural and intergenerational exchange of ideas through making. With our approach, we challenge the traditional definition of *community outreach* performed by academic makerspaces. Existing academic makerspaces often perform community engagement by only offering hourly, one-time workshops or by having community members provide a problem that is then used by students as a project assignment. These approaches position community members as occasional visitors and non-equal contributors, which not only conflict with the core values of co-creation but also limit the makerspaces' impact on connecting the universities and the communities. C@L explored an alternative approach in which we invited community members as long-term and equal co-makers into the academic makerspaces.

To enable long-term community partnership, C@L focused on creating the *continuity of people* through skill sharing, proposing projects, and mentoring projects. In Fall 2019, C@L launched a community hackerspace that provided an open space where students and community members from neighborhoods around Cornell Tech came to ask questions, learn design and making skills, and execute their own projects using the latest digital fabrication tools. In addition, C@L hosted a 15-week workshop for students and community members to co-design solutions for projects initiated by community members. Moreover, C@L structured programs in which community members mentored projects conducted by Cornell Tech students and other community members (Figure 1).

In this article, we showcase two sets of collaborations that illustrate the continuity of people through co-making. The first was a one-to-one collaboration between a Ph.D. student and a community mentor evaluating, improving, and ultimately finalizing a previously co-designed artifact. The second collaboration was a design studio-like weekly session where community members and students collectively ideate, prototype, and build low-cost weaving artifacts. Through these collaborations, we present how academic makerspaces can function as a hub that connects community members and partner organizations with the campus community in a long-term relationship.

## BACKGROUND

### Makerspace in Public Life

The Maker Movement has recognized the broadening impact of makerspaces in public life, such as serving as social hubs, supporting wellbeing, connecting excluded communities [1]. Academic makerspaces have been mostly engaging communities through K-12 education and continued education [2]. Our work examined how academic makerspaces can engage more diverse communities more closely through non-educational approaches.

### Co-Making

Co-creation has received increasing attention and usage across public policy, art, design, and computing [3]. Each field grew its vocabulary (participatory design, co-production, etc.) to denote the core value of co-creation: the dissemination of control over production and the sharing of expertise [3, 4]. In this article, we use the word *co-making* to capture this notion in the maker movement. We define that, specifically to academic makerspaces, co-making refers to the share of design participation, decision, and expertise between academic and community members.

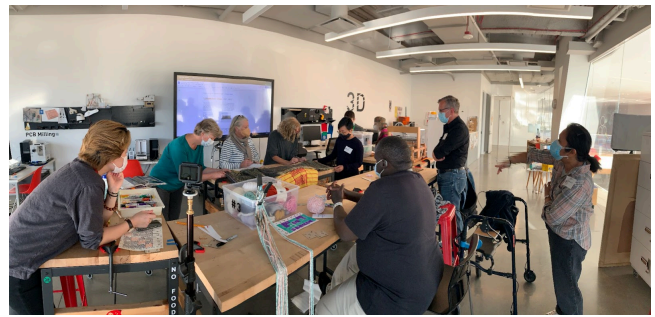

*Fig 1. Community members and students worked together in C@L.*

## COLLABORATIONS

We detailed how we structured co-making to achieve the continuity of people.

*Word Tiles*

Word Tiles is an artifact first co-designed by residents of a long-term care facility center and students through semester-long credited workshops. After the semester ended and students graduated, a community mentor and a Ph.D. student continued the development efforts.

The design goal for Word Tiles is to reconnect the residents in the long-term care facility with each other after COVID-caused social isolation. Community members decided that the artifact should be a board game in service of their fondness for board games. Students created an initial prototype of the game in MakerLAB. Following this initial step, a Ph.D. student and a community mentor deployed the artifact and evaluated its impact through qualitative interviews and contextual inquiries. Through iteration, the artifact was made stand-alone and utilized in the care facility.

The outcome of Word Tiles is threefold. The first layer is *fulfilling the design goal*. The community members were able to define a problem that they cared about and design solutions for themselves. This enabled the artifact's forms, such as its tangibility, font size, and playing method, to be suitable for the specific use environment of the specific users. The second layer is *achieving longevity*. The continuous involvement of community members ensured that the Word Tiles' development was independent of the length of the workshop series or the participating students graduating: students are more temporary, but community members typically stay longer, especially with the projects they are deeply involved in. The last and the most important is *empowering community members* through participation. The residents who participated were proud that their design circulated the care facility and was used by other residents.

*Community Loom*

Community Loom was an idea initiated by the community mentor who guided in the Word Tiles project. C@L brought a team together that consisted of the long-term care facility residents, seniors, art therapists, textile experts, artists, design hobbyists, and students to bring this idea to life. C@L aimed to create the atmosphere of a design studio: team members gathered weekly in the maker space, learned about the overall design goal, sh ared skills, performed rapid prototyping, and informally conducted design crits.

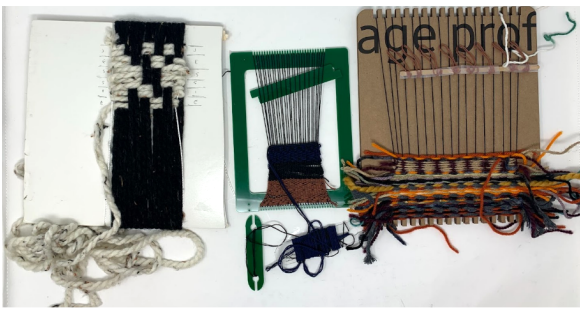

**Fig.2   Artifacts prototyped through Community Loom.**

The design goal of Community Loom highlighted the exchange of visions. The community mentor envisioned an easy-to-hold and straightforward weaving of artifacts for residents to use during art therapies. The MakerLAB envisioned an exhibition, pop-up, or installment that would be set up at or traveling through various points in the community. Although the forms of these two visions suggest distinctive forms, they are guided by the same concept of bringing communities together through weaving.

The collaboration that formed to create the Community Loom is arguably more important than the artifact outcomes. The distinctiveness of the two visions provided rich possibilities for the diverse team members to find their interests. For instance, the textile experts discussed how to simplify the weaving procedures for easy onboarding and how to embed unique identifiers to weave patterns. The art therapists and residents partnered with artists and design hobbyists to play with prototypes and iterate on various loom designs. Students were interested in engineering and designing digital interaction, including AR-powered weave pattern reading.

## DISCUSSION AND FUTURE WORK

Word Tiles and Community Loom showed that co-making through structured, one-to-one collaboration and unstructured design studio can enable community members to form long-term connections with academic makerspaces. We further demonstrated that academic makerspaces could be the anchor for universities to engage with diverse community groups and the hub for members in the community to socialize and collaborate. We hope our current progress on C@L could elicit new ways and new ideas of how academic makerspaces position themselves in the university-community connection as well as a starting point to engage with communities generally marginalized in academia.

## ACKNOWLEDGEMENT

We thank PiTech Doctoral Fellowship and Cornell Tech for supporting this work. We also thank the participation from NYC H+H/Coler on Roosevelt Island and CBN Senior Center.